\documentclass[prl,twocolumn,aps,byrevtex,showpacs,tightenlines]{revtex4}
\usepackage{amsmath}
\usepackage{amsfonts}
\usepackage{epsfig}

\begin{document}
\title{Equivalent efficiency of a simulated photon-number detector}
\author{Kae Nemoto}
\author{Samuel L.\ Braunstein}
\affiliation{Informatics, Bangor University, Bangor LL57 1UT, UK}
\date{\today}

\begin{abstract} 
Homodyne detection is considered as a way to improve the efficiency 
of communication near the single-photon level. The current lack of 
commercially available {\it infrared\/} photon-number detectors 
significantly reduces the mutual information accessible in such a 
communication channel. We consider simulating direct detection via homodyne 
detection. We find that our particular simulated direct detection strategy 
could provide limited improvement in the classical information transfer.
However, we argue that homodyne detectors (and a polynomial number of 
linear optical elements) cannot simulate photocounters arbitrarily 
well, since otherwise the exponential gap between quantum and classical 
computers would vanish.
\end{abstract}
\pacs{03.67.Lx, 42.50.Dv, 89.70.+c}
\maketitle

The fundamental limitations to classical communication in optical channels 
are due to the quantum nature of the signals being transmitted. These 
limitations have been well understood for {\it ideal\/} optical 
communication channels \cite{Yuen93,Caves94}. The capacity of a 
communication channel is defined to be the maximum mutual information 
(optimized over the choice of source alphabet used by the sender and the 
detection strategy used by the receiver) across the communication channel 
with respect to some physical channel constraint --- such as the mean 
energy throughput. This characterization of a communication channel 
is important because Shannon's noisy coding theorem proves that any 
attempt at communication beyond this capacity necessarily fails due 
to unrecoverable errors \cite{Shannon63}. A corollary to this theorem 
states that even for a non-optimal source alphabet or detection strategy, 
the mutual information of the communication channel is (asymptotically) 
achievable using error correction \cite{Shannon63,Gallager68}. 

For a single-mode optical communication channel the optimal capacity, 
under a mean energy constraint, is achieved with a source alphabet of 
photon-number states and ideal photon-number detectors 
\cite{Yuen93,Caves94}. In this ideal case the orthogonality of the 
signals and hence their perfect distinguishability makes error 
correction unnecessary. Unfortunately, however, {\it such ideal operation 
is currently impractical}, since neither ideal photon-number state 
preparation nor ideal photon-number detection is achievable.

The detection of weak signals (few photons) is especially difficult at
communications wavelengths ($1.3 - 1.55 \mu$m). Ideally, we would wish
to achieve this by simply counting the photons. Now the process of 
photo-counting is often synonymous with using an avalanching device with 
saturated gain, since each photon produces a strong and standard signal 
at the output. Unfortunately, it is a technological fact that both at 
infrared and optical frequencies, the best avalanche photodiodes
never have as high a quantum efficiency as the best available linear 
detectors (having linear gain, such as PIN-photodiodes). In fact, currently, 
no {\it commercial\/} photocounters are available at communications 
frequencies. Thus, at these frequencies PIN-photodiodes are routinely
used despite their high dark count rates. Partly because of this, the 
traditional solution at communications wavelengths has been to encode 
signals on intense (many photon) pulses.

At communication frequencies, photon counting has been achieved  
with InGaAs or Ge avalanche photodiodes operating in so-called Geiger
mode \cite{Zbinden98,Stucki01}. Due to the high dark count rate, 
performance of these detectors as photon counter is very low. The best 
efficiency reported is around 20\% at 1.3 $\mu$m with the optimal 
temperature 77K \cite{Zbinden98}, and is around 10\% at 1.5$\mu$m with 
the optimal temperature 213K
\cite{Stucki01}.

In this paper we consider an alternate encoding and detection strategy
which is suitable for truly weak signals and current technological
limitatons. The basic idea is to {\it simulate\/} direct detection via 
a dual-homodyne scheme. Because strong local oscillators are continuously 
producing a strong output photocurrent, even high dark count rate detectors 
like PIN photodiodes, which have the highest quantum efficiencies, may 
be used. For example, at communication frequencies such devices can have 
efficiencies appraoching 90\% \cite{Silberhorn01} and even 98\% at optical 
frequencies \cite{Polzik92}. We determine the mutual information for 
inefficient direct detectors and compare it to that of efficient homodyne 
detectors for a source alphabet preferring direct detection strategies. 
In doing so we are able to compute an equivalent efficiency for our 
simulated photon-number detectors for communication purposes.

As we shall see, there is a communication penalty for simulating direct 
detection in this way. Notwithstanding this penalty, the reduced 
signal-to-noise due to high dark counts suggests that our indirect 
detection strategy is worth consideration.

Let us first consider two random variables $A$ and $B$, with individual 
values labelled $a$ and $b$ respectively and a joint probability of 
$P_{a,b}$. The mutual information between these variables is defined by
\begin{equation} \label{mi}
{\cal I}(A:B)= \sum_{a,b} P_{a,b} \log \left(\frac{P_{a,b} }{P_{a}P_{b}}
\right) \;,
\end{equation}
and is, in some sense, a measure of the information content that is common
to both variables. This quantity of mutual information is important
in communication theory, because it can be used to quantify the information
content that a receiver, observing variable $B$, learns about the
sender's message represented by variable $A$.

\begin{figure}[hbt]
\epsfig{figure=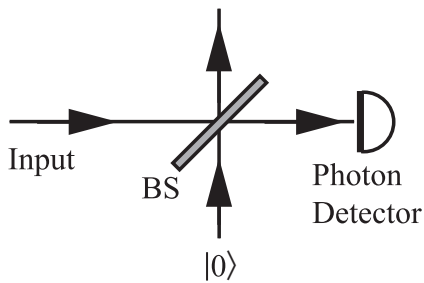,width=55mm}
\caption{The non-perfect measurement scheme: this models the non-perfect
direct photon number detection with a perfect photon counter and 
the beam-splitter (BS) which determines the finite efficiency as 
$\eta^2$.}
\label{pd1}
\end{figure}

The optimal capacity of an ideal bosonic communication channel with a
mean energy constraint is achievable with a mean-channel state that is 
thermal and is calculated from Eq.~(\ref{mi}), yielding \cite{Yuen93} 
\begin{equation} \label{max}
{\cal I}^{pd}(A:B)=(1+\bar{n})\log{(1+\bar{n})}
-\bar{n}\log{\bar{n}}\;,
\end{equation}
where $\bar{n}$ is the mean photon number of the thermal state
\begin{equation} \label{thermal}
P_n=\frac{1}{1+ \bar{n}} \Big(\frac{\bar{n}}{1+\bar{n}} \Big)^n \;.
\end{equation}
We model loss by introducing extra beam-splitters into the channel or in
front of the detector, discarding photons in the unused port. A schematic 
representation of this is shown in Fig.~\ref{pd1}. The parameter $\eta$ is 
the {\it amplitude\/} efficiency corresponding to the amplitude-reflection
coefficient of the beam-splitter, so $\eta^2$ represents the quantum 
efficiency of the overall detector.  

\begin{figure}[hbt]
\epsfig{figure=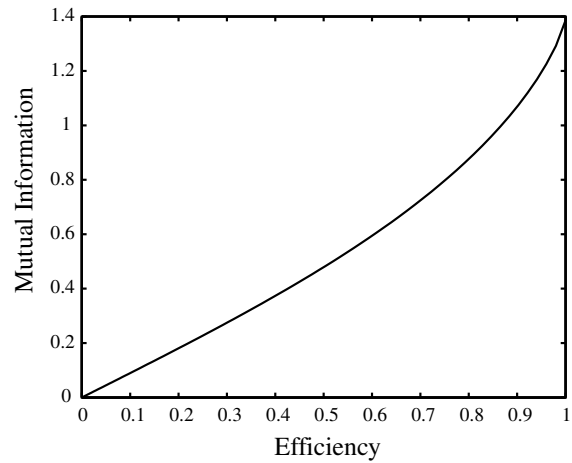,width=75mm}
\caption{The mutual information versus quantum efficiency $\eta^2$ of 
a non-perfect measurement scheme is plotted for a mean photon number of one.  
This curve shows that a slight reduction below perfect efficiency causes
a significant reduction in mutual information.}
\label{finite}
\end{figure}

Assume that the input signal is characterized by a mean-channel state
that is thermal (\ref{thermal}), with vacuum entering the second input of 
the beam-splitter, then the probability distribution for detecting
$m$ photons with our model of an inefficient detector is is given by
\begin{equation}
P_m = \frac{1}{1+\eta^2 \bar{n}}\Big( \frac{\eta^2 \bar{n}}{1+\eta^2 \bar{n}}
\Big)^m\;.
\end{equation}
This corresponds to a Poisson distribution with reduced mean number of
detected photons, down to $\eta^2 \bar{n}$. The mutual information for 
a source alphabet of number states and inefficient detection is then
simply
\begin{eqnarray} \label{inf_dir}
{\cal I}^{\eta}(A:B) &=& \log{\Big[ (1-\eta^2)^{\bar{n}}(1+\eta^2
\bar{n})\Big]} \nonumber\\ 
&&+ \eta^2\bar{n} \log{\Big[ \frac{(1+\eta^2\bar{n})}{(1-\eta^2)
\bar{n}}\Big]} \nonumber\\
&&+ \frac{1}{1+\bar{n}}\sum^{\infty}_{n=0}\sum^n_{m=0} \eta^{2m}
(1-\eta^2)^{(n-m)}\nonumber\\
&&\phantom{+} \times  \Big( \frac{\bar{n}}{1+\bar{n}} \Big)^n
    \left( \! \begin{array}{c}
    n\\ 
    m
    \end{array}\!\right)
    \log{\left( \!\begin{array}{c}
    n\\ 
    m
    \end{array}\!\right)
  }\;.
\end{eqnarray}
We observe in Fig.~\ref{finite} that the mutual information decreases for 
finite loss of photons. In particular a small amount of loss away from
perfect detection results in a significant decrease in the mutual 
information, while the loss of mutual information is almost linear 
in the low efficiency regime $\sim 0.5$. This implies that a small 
improvement in the photon detection efficiency of current technology 
cannot be expected to bring a significant increase in the information
throughput.

\begin{figure}[ht]
\epsfig{figure=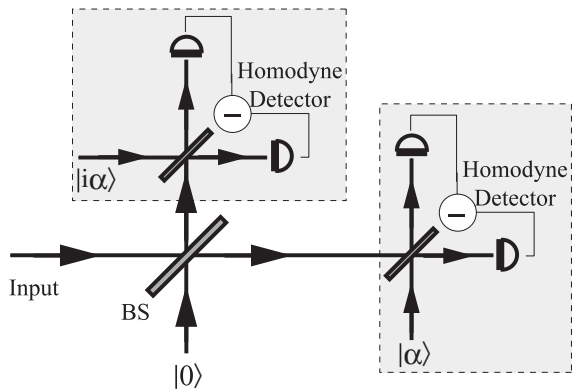,width=75mm}
\caption{Dual homodyne measurement strategy: each gray box shows an
individual homodyne detector to measure a different quadrature-phase 
amplitude of the input signal.}
\label{qnd}
\end{figure}

We now consider near-ideal homodyne measurements in comparison to non-ideal
photon counting. As we have discussed, such measurements can achieve very 
high efficiencies because homodyne detectors may be operated without
regard to dark current. Hence as a first approximation we will treat the 
homodyne measurements as ideal. Our detection strategy is based on dual 
homodyne detection, which can simultaneously detect both quadrature-phase 
amplitudes. A schematic representation of dual homodyne detection is shown
in Fig.~\ref{qnd}.  

Suppose that a number state is sent down the channel as an input signal and
vacuum enters the first beam splitter as its second mode (as shown in
Fig.~\ref{qnd}). This specifies an input $A$ as 
$|n\rangle \otimes |0\rangle$, or simply denoted $|n0\rangle_A$. Similarly 
the output $B$ is collapsed into eigenstates of quadrature-phase amplitudes
which are analogs of position and momentum and we denote by $|XP\rangle_B$ 
states. The probability of measuring $X$ and $P$ at the output $B$, given
an input $A$ specified by $|n0\rangle_A$, is given by the square of 
the inner product $_A \langle n0| XP \rangle_B$. Taking $|k\rangle$ as
number states, this probability may be written
\begin{eqnarray}
\label{conditional}
P_{X,P|n}&=&\Big|\sum^n_{k=0}  \left( \begin{array}{c}  
				n\\ 
				k
			 \end{array}\right)^{1/2}
			 \sqrt{2}^{-n} \langle X| k \rangle
			 \langle P| n-k \rangle \Big|^2 \nonumber \\
&=&\frac{1}{2^{2n} \pi n!} e^{X^2+P^2} \Big| \Big( \frac{\partial}
{\partial X}
-i \frac{\partial}{\partial P}\Big)^n e^{-(X^2+P^2)}\Big|^2\;, \nonumber \\
&=&\frac{1}{\pi n!}e^{uv}\Big|
\Big(2\frac{\partial}{\partial v}\Big)^n
e^{-uv}\Big|^2_{(u,v)=(X-iP,X+iP)}\;.
\end{eqnarray}
Clearly, this probability $P_{(X,P)|n}$ is dependent only on the product
$uv$ ($=X^2+P^2$), hence we change the variables to the polar coordinates,
$(I^{1/2},\theta)$ where the intensity $I\equiv X^2+P^2$, yielding   
\begin{equation}
P_{(X,P)|n}= \frac{1}{\pi n!} I^n e^{-I}\;.
\end{equation}
Finally, integrating this over all angles $\theta$ gives
\begin{equation}
\label{cond}
P_{I|n}	= \int^{2\pi}_0 d\theta P_{(I, \theta)|n}
	= \frac{1}{n!} I^n e^{-I}\;.
\end{equation}

Now the quantity $P_{I|n}$ is a conditional probability for a given 
input photon number $n$. From it we can compute the unconditioned
probability averaged over the mean-channel thermal state with a
a mean photon number $\bar{n}$. The resulting distribution is
\begin{equation}
\label{uncond}
P_I = \sum^\infty_{n=0} P_{I|n} P_{\bar{n}}
= \frac{1}{1+\bar{n}} e^{-\frac{I}{1+\bar{n}}}\;.
\end{equation}
These probabilities from Eqs.~(\ref{cond}) and~(\ref{uncond}) allow us 
to calculate the mutual information between sender and receiver for this 
dual homodyne detection scheme. It is given by
\begin{eqnarray}
\label{inf_qnd}
{\cal I}^{hd}(A:B)
&=&(1+\bar{n})\log{(1+\bar{n})} - \gamma \bar{n} \nonumber \\
&&-\sum^{\infty}_{n=1}
(\frac{\bar{n}}{1+\bar{n}})^n \log{n} \;,
\end{eqnarray}
where $\gamma=0.5772\ldots$ is Euler's constant. We note here that we could
have replaced the dual homodyne detection by hetrodyne measurement 
\cite{Caves85,Caves86}.

\begin{figure}[ht]
\epsfig{figure=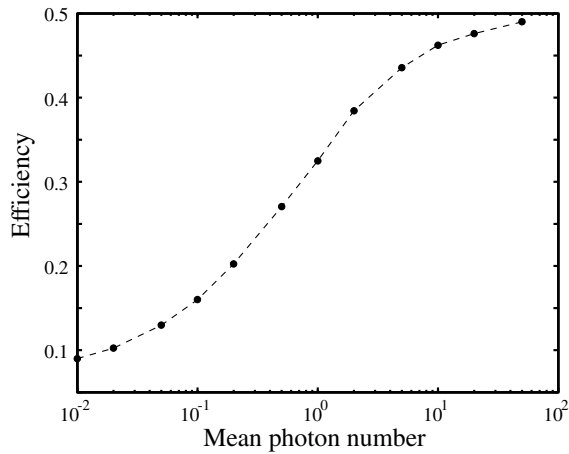,width=75mm}
\caption{The equivalent efficiency ${\eta^*}^2$ determined by inverting
Eq.~(\ref{equiv-eff}); shown with a semi-log scale in the mean photon number.}
\label{efficiency}
\end{figure}

We now introduce a measure of efficiency for the dual homodyne measurement.
There are a number of possible ways to evaluate efficiency of detection 
schemes. The measure of efficiency that we introduce here is based on a
comparison with the direct detection scheme with finite efficiency.    
In particular, we will equate the mutual informations achievable in each
of these two schemes with an alphabet chosen to prefer direct detection.
The choice of finite eifficiency ${\eta^*}^2$ in the direct detection scheme 
for which this equivalence holds is dubbed by us the {\it equivalent 
efficiency\/} of the dual homodyne detection scheme (at least for the 
purposes of classical communication as analyzed here). Thus, the
equivalent efficiency may be determined by inverting the relation
\begin{equation} \label{equiv-eff}
I^{\eta^*}(\bar{n})={\cal I}^{hd}(\bar{n})\;,
\end{equation}
for the efficiency ${\eta^*}^2$. (In other words, the photon detection 
efficiency is chosen to be that efficiency for which the mutual information
obtained by direct detection is exactly that given by dual homodyne 
detection.) This efficiency is obviously dependent on the mean photon 
number $\bar{n}$ for the source alphabet. For instance, the equivalent 
efficiency of the dual homodyne measurement for a mean-photon number of 
one is just ${\eta^*_1}^2\simeq 0.327$. More general cases are shown
on the graph of equivalent efficiency versus mean phont number in
Fig.~\ref{efficiency}. The region $0.1 \alt \bar{n} \alt 1$ is the most
sensitive to the mean photon number in the growth of the equivalent
efficiency ${\eta^*}^2$. For smaller mean photon numbers, the mutual
information $I^{hd}(\bar{n})$ of Eq.~(\ref{inf_qnd}) does not gain much 
by small increases in the mean photon number. Similarly, for $\bar{n}\agt 1$
no significant gain in equivalent efficiency is obtained for small changes
in the mean photon number of the source alphabet.

For an input alphabet of photon-number states, it is clear that
schemes based upon homodyne measurement cannot be expected to perform
as well as ideal direct photon detection. Nonetheless, such ideal direct 
photodetectors are not currently technologically realistic, especially
at communications wavelengths. By contrast, since homodyne detectors may 
be operated without regard to dark current a significantly higher 
quantum efficiency is readily available for them. We have found that
replacing inefficient direct detectors with homodyne-based simulated 
direct detectors can yield reasonable improvements, even near the 
single-photon level of operation. In this paper, we have shown that this 
improvement is theoretically possible for the purposes of classical 
communication through a single-mode bosonic channel. Our analysis used 
a communication alphabet of photon-number eigenstates, which were thus 
optimized for (ideal) direct detection schemes. This choice is strongly
prejudiced towards direct detection and against our homodyne scheme. Thus,
our estimate of equivalent efficiency is likely to be an underestimate
of the performance of homodyne-based schemes in general. If instead, we 
had optimized the input alphabet for homodyne detection, we would have
seen a significant improvement in capacity \cite{Caves94}. However, since 
this capacity is very likely to be larger than that possible for the 
imperfect direct detection schemes our strategy for computing a figure
of `equivalent efficiency' would not be applicable.

Another aspect that is missing from the analysis given here, is a 
detailed consideration of error correction coding schemes that would be 
required to achieve the performance promised by Shannon's measure of
communication throughput, namely, the mutual information. In general, 
more complicated encoding will be required to achieve the information 
transfer given by this measure. 

Finally, it remains to be considered whether the approach studied here
really has any applicability to either quantum communication or 
computation. The maximum 50\% equivalent efficiency of the simulated 
photon detection here might rule out these possibilities for detecting 
quantum information represented within discrete photonic Hilbert spaces.
Thus, if we hope to use these ideas beyond classical communication this 
low efficiency implies that discrete Hilbert spaces will need to be 
abandoned. This suggests using homodyne detectors within some kind of
continuous quantum variable scenario. For such variables a generalized
Gottesman-Knill theorem has been derived \cite{Bartlett01}. This theorem
states, under relatively mild assumptions, that quantum computational 
circuits consisting of gate operations made from quadratic Hamiltonians and 
homodyne-based measurements can be efficiently simulated on a classical 
computer. Further, including direct photodetection is sufficient to 
provide these circuits with the capability to perform universal quantum 
computations \cite{Gottesman}. This observation suggests that our 
particular simulation strategy cannot be improved arbitrarily. 
For if homodyne-based measurements 
(and linear optics) could come arbitrarily close to simulating 
photocounting with only a polynomial number of components then a universal 
quantum computer could be simulated classically in polynomial time by
this above theorem, which would imply $\text{BQP}=\text{BPP}$.
($\text{BPP}$ is the class of problems that can be solved using
randomized algorithms in polynomial time, while $\text{BQP}$ is the
class of all computational problems which can be solved efficiently on
a quantum computer.)  This 
outrageously unlikely outcome suggests that simulating direct detection 
using homodyne detectors must have limited efficiency.

\vskip 0.1truein

This work is funded by the Research Foundation for Opto-Science and 
Technology.  

\end{document}